# Sign reversal in the exchange bias and the collapse of hysteresis width across the magnetic compensation temperature in a single crystal of $Nd_{0.75}Ho_{0.25}Al_2$

Prasanna D. Kulkarni, A. Thamizhavel, V. C. Rakhecha, A. K. Nigam, P. L. Paulose, S. Ramakrishnan and A. K. Grover

Department of Condensed Matter Physics and Materials Science,

Tata Institute of Fundamental Research, Colaba, Mumbai, India 400005.

Abstract

In the admixed $Nd_{0.75}Ho_{0.25}Al_2$ system, magnetic moments of Nd and Ho occupying the same crystallographic site are antiferromagnetically coupled and the chosen stoichiometry displays a magnetic compensation behavior ($T_{comp} \approx 24$ K) in all orientations. In the vicinity of $T_{comp}$, the conduction electron polarization (CEP) assumes the role of a soft ferromagnet exchange coupled to a pseudo-antiferromagnet comprising Nd/Ho moments, resulting in an asymmetry in the hysteretic (*M-H*) loop, i.e., the notion of an exchange bias field ($H_{exch}$). Across $T_{comp}$, the CEP contribution reverses sign, and in consonance, the asymmetry in the *M-H* loop also undergoes a phase reversal. Interestingly, the width of the *M-H* loop shows a divergence, followed by a collapse on approaching $T_{comp}$ from either end. The observed behavior confirms a long standing prediction based on a phenomenological model for ferrimagnetic systems. The field induced changes across $T_{comp}$ leave an imprint of a *quasi*-phase transition in the heat capacity data. Magneto-resistance ($\Delta R / R$ vs $T$) has an oscillatory response, in which the changes across $T_c$ and $T_{comp}$ can be recognized.

**Introduction**

The basic physics of conventional ferrimagnets encompassing two magnetic sublattices with unequal magnetic moments antiferromagnetically coupled to each other is well known [1,2]. On warming up a ferrimagnet in a moderate applied field ($H \sim 10$ kOe), if the larger of the two moments decreases faster than the smaller one, a circumstance can arise that the net magnetization attains a minimum value in the ordered state, thereafter, the contribution to the magnetization from the smaller moment dominates and the temperature dependent magnetization curve ($M(T)$) displays a maximum prior to crossing over to the paramagnetic state. The notion of magnetic compensation in ferrimagnets (i.e., minimum/maximum, followed by maximum/minimum in a $M(T)$ curve on warming up/cooling down) is a common occurrence [1-3]. Another hallmark of compensation phenomenon in ferrimagnets is that the $M(T)$ curve, measured in a field less than $\sim 100$ Oe, crosses from positive to (metastable) negative values at a temperature designated as $T_{comp}$. Adachi, Ino and Miwa [4] rekindled the interest in exploration of magnetic compensation in a particular variety of ferromagnetic Samarium systems, which get easily driven to the 'net-zero' magnetization limit on substitution of $Sm^{3+}$ ions with a few percent of $Gd^{3+}$ [5] or $Nd^{3+}$ ions [6,7], depending on the pristine state of the Sm magnetization in a given host to be orbital surplus or spin surplus [4,6,7]. In these Samarium systems, there are no two magnetic sublattices with unequal moments, instead the (nearly equal) orbital ($<L_z>$) and spin ($2<S_z>$) contributions to the magnetization of the $4f$-$Sm^{3+}$ ion ($<\mu_z^{Sm}> = -\mu_B <L_z + 2S_z>$, where $<L_z> = -30/7$ and $<2S_z> = 25/7$ for free $Sm^{3+}$ in the ground multiplet), compete with each other in the presence of a small, but, important contribution from the conduction electron polarization (CEP) [6]. The net magnetization in a realistic Sm system in presence of crystalline electric field effects is typically even smaller ($< 0.5$ $\mu_B$ / formula unit ($f.u.$)) [8] than the free ion value of $5/7$ $\mu_B$; it is either orbital surplus or spin surplus and can be taken towards the zero magnetization limit by substitution of a very small fraction of $Sm^{3+}$ ions by (spin only) $Gd^{3+}$ or (orbital-surplus) $Nd^{3+}$ ions having magnetic moments much larger than that due to Samarium. The rationale for such a behaviour in admixed (ferromagnetic) Sm based alloys is that the indirect RKKY exchange interaction, mediating via the CEP, keeps the "spins" of all the 4$f$-rare earth (RE) ions (e.g., $Sm^{3+}$ and $Gd^{3+}/Nd^{3+}$) ferromagnetically aligned [9], thereby, resulting in an antiferromagnetic (ferromagnetic) coupling between "magnetic moments" of Sm and Gd (Nd) ions. An interesting aspect reported in the doped $Sm_{1-x}Gd_xAl_2$ series (where $x$ = 0.01 and 0.02 and $SmAl_2$ has $T_c$ = 125 K and $\mu$ = 0.2 $\mu_B$ / $f.u.$) is the progressively enhancing fingerprint of the field-induced reversal in the orientation of magnetic moment of the Sm ions w.r.t. the applied field across $T_{comp}$ in the specific heat data [10]. Such an observation calls for association of notion of field-induced phase transition across $T_{comp}$ in the admixed Sm based system. Magnetic compensation, *per se*, had been studied in 1960s in several admixed $R_{1-x}R'_xAl_2$ alloys [11,12], where $R$ and $R'$ belong to the different halves of 4$f$ RE series, but phase transition aspect across $T_{comp}$ had not been explored.

We have now performed detailed magnetization, specific heat and magneto-resistivity measurements in a single crystal of $Nd_{0.75}Ho_{0.25}Al_2$, this stoichiometry in the polycrystalline form had been shown to imbibe a magnetic compensation feature by Swift and Wallace [12]. An exciting new finding is that as the competing contributions from magnetic moments of Nd and Ho moments to the 'net-magnetization' reverse their orientations across $T_{comp}$, the contribution from the CEP also reverses its sign over a narrow temperature interval, and it vividly shows up in the temperature variation of the asymmetric shift of the magnetization hysteresis loop (the so called 'exchange bias field', $H_{exch}$), which also reverses its phase across $T_{comp}$. Another significant observation is the concurrent sudden collapse on approaching $T_{comp}$ in the half-width of the hysteresis loop, which measures the 'effective coercive field' in the sample. The field-induced magnetic reorientation(s) across $T_{comp}$ also leave their imprint in the specific heat and the magneto-resistivity data.

**Experimental details**

The stoichiometry $Nd_{0.75}Ho_{0.25}Al_2$, made up of two ferromagnetic compounds, $NdAl_2$ ($T_c \sim 70$ K, $\mu/Nd^{3+} \sim 2.5$ $\mu_B$) and $HoAl_2$ ($T_c \sim 33$ K, $\mu/Ho^{3+} \sim 8.1$ $\mu_B$), is estimated to be close to the nominal zero magnetization limit from the plot of magnetization values in $Nd_{1-x}Ho_xAl_2$ series [cf. Fig. 4, Ref. 12]. A single crystal was pulled from the polycrystalline melt of $Nd_{0.75}Ho_{0.25}Al_2$ by Czochralski method, using the tetra-arc configuration (Techno Search Corp. Japan, Model: TCA 4-5), at a linear speed of 10 mm/hr for ~ 8 hrs. Its Laue x-ray diffraction pattern could be reconciled to the cubic Laves phase (*C15*) structure. Three platelet shaped crystals were cut by the spark erosion, such that the directions perpendicular to the planes of the platelets were along the three principal crystallographic axes, viz., [100], [110] and [111]. The isofield DC magnetization measurements were performed on all the three crystal pieces using a commercial SQUID magnetometer (Quantum Design (QD) Inc., USA, Model MPMS 7). The isothermal magnetic hysteresis loops were recorded on a crystal with $H \parallel$ [100] using a SQUID-Vibrating Sample Magnetometer (QD Inc., USA, Model SVSM). The electrical resistance and the heat capacity measurements were carried out on the same crystal piece using a Physical Property Measurement System (QD Inc., USA, Model PPMS).

**Results and discussion**

Figures 1(a) and 1(b) show the temperature dependences of the dc magnetization (*M-T* curves) measured while cooling in nominal zero field ($H \sim 1$ Oe) and 5 kOe, respectively in the three crystals of $Nd_{0.75}Ho_{0.25}Al_2$, with the field oriented in [100], [110] and [111] crystallographic directions. The magnetic ordering temperature ($T_c$) has been marked at ~ 70 K in Figs. 1(a) and 1(b), it is close to the nominal ordering temperature of the pure $NdAl_2$ alloy. While cooling the three crystals below $T_c$ in nominal zero field, the net positive magnetization values are observed till ~ 24 K (cf. Fig. 1(a)) and it can be noted that [100] is the easy direction. At ~ 24 K, the three *M-T* curves in Fig. 1(a) cross the $M = 0$ axis towards the metastable negative values, thereby elucidating the isotropy and intrinsic character of the magnetic compensation at the given stoichiometry. A schematic drawn in Fig. 1(a) shows the notional relative alignments of the magnetic moments of Nd and Gd and that of CEP w.r.t. the applied field at all temperatures at low fields. The $M(T)$ curves at 5 kOe in Fig. 1(b) show a sharp turnaround at ~ 24 K. Interestingly, the $M(T)$ curve for [110] in 5 kOe lies a little above that for [100], a situation unlike the one in Fig. 1(a) at H ~ 1 Oe. The turnaround in $M(T)$ curves in Fig. 1(b) reflects the field-induced reversal in the orientations of magnetic moments of Nd/Ho and that of CEP.

The relative orientations of magnetic moments contributing to the magnetization signal can also be explored in the warm up of the remanent ($M_{rem}$) signal, as depicted in the inset panel of Fig. 1(a). The remanence at 5 K was obtained by cooling the crystal in a magnetic field ($H \parallel$ [100]) of 50 kOe from the paramagnetic state, the field was then ramped down and magnetization signal was measured in 50 Oe in the warm up mode. In the inset panel of Fig. 1(a), the warm up of $M_{rem}$ is displayed along with the field cool $M(T)$ curve in the same field. The mirror reflection characteristic and the intersection of the two curves at $T_{comp}$ of 24 K endorse the rigidity in the antiferromagnetic coupling between the magnetic moments of Ho and Nd. The $M_{rem}(T)$ curve can be seen to eventually merge into the cool down $M(T)$ curve while crossing over to the paramagnetic state.

To further characterize the field induced reversal process, we show in Figs. 2(a) to 2(c) the highlights of the magnetization, heat capacity and magnetoresistance data obtained in fields ≥ 10 kOe ($H \parallel$ [100]). The $M(T)$ curve in 20 kOe in Fig. 2(b) shows a sharp turnaround feature at 29 K, followed by another shallow dip near 20 K. It is known that the easy axis of magnetization in pure $HoAl_2$ changes from [100] to [110] above 20 kOe at ~ 20 K. We, therefore, reckon that after the Nd/Ho moments reverse their orientations w.r.t. field applied along [100] direction at 29 K, another (partial) orientation occurs as the temperature lowers towards 20 K, reflecting the tendency of Ho moments to shift from [100] to [110] direction locally. The $M(T)$ curve at 10 kOe in Fig. 2(a) has a broad valley extending from 22 K to 26 K, as compared to sharp features of *M(T)* curves in 5 kOe (cf. Fig. 1(b)) and 20 kOe (cf. Fig. 2(a)). In the main panel of Fig. 2(b), where we present a comparison of the heat capacity ($C_p$) data in a field of 20 kOe with that in zero field, one can immediately notice the surfacing of a sharp anomalous peak near 29 K. The inset (i) in Fig. 2(b) shows a blow up of the $C_p$ versus *T* plot centered near 29 K in H = 20 kOe. The field induced enhancement $\Delta C_p$ can be ascertained from this plot. The inset (ii) in Fig. 2(b) shows a plot of $\Delta C_p$ versus *H* in the $Nd_{0.75}Ho_{0.25}Al_2$ crystal from H = 10 kOe to 50 kOe. Qualitatively, this behavior appears analogous to that reported in the polycrystalline $Sm_{1-x}Gd_xAl_2$ alloys (*x* = 0.01 and 0.02), though the $\Delta C_p$ values in the single crystal under study are an order of magnitude higher.

An inset panel in Fig. 2(c) displays the resistance versus temperature plot, *R(T)*, in the single crystal of $Nd_{0.75}Ho_{0.25}Al_2$ for *H* || [100] and electrical current transmitted normal to the field. A sharp knee like feature in the *R(T)* curve at 70 K reflects the onset of drop in spin-flip contribution to the resistivity [13] at the onset of the magnetic ordering in the sample. We had recorded *R(T)* curves in different applied fields to explore magnetoresistance. The main panel of Fig. 2(c) show the temperature variation of magneto-resistance, $\Delta R(H) / R(0)$, where $\Delta R(H) = R(10\ kOe) - R(0)$. The oscillatory behavior of $\Delta R(H)/R$ versus *T* in $Nd_{0.75}Ho_{0.25}Al_2$ crystal appears perplexing. Note first that in the paramagnetic state $\Delta R/R$ is negligible, however, as the critical slow down is stabilized by the applied field on approaching the $T_c$ (≈ 70 K), the onset of drop in spin-flip scattering sets in a little above $T_c$ (compare its $C_p$ versus *T* plot in H = 20 kOe in Fig. 2(b)) and $\Delta R/R$ shows a negative peak centered around 70 K. Interestingly, $\Delta R/R$ versus *T* curve first crosses from negative to positive values at ~ 55 K, which corresponds to temperature of broad maximum in *H(T)* curve in Fig. 2(a). Thereafter, the curve once again crosses to the negative values at the turnaround temperature of 29 K. $\Delta R/R$ versus *T* curve changes direction for the last time at $T_{comp}$ of 24 K and returns towards positive value at *T* about 19 K.

Figures 3 and 4 summarize the key results of the magnetization hysteresis measurements in $Nd_{0.75}Ho_{0.25}Al_2$ crystal for $H \parallel [100]$. We had recorded the hysteresis (*M-H*) loops at different temperatures: The single crystal was cooled to a given temperature in a field of 5 kOe from the paramagnetic state and the *M-H* loop was then traced by cycling the field between ± 5 kOe. The temperature intervals were chosen to get fairly detailed information on the hysteretic behavior very close to the compensation temperature (≈ 24 K). Figure 3 shows portions of the *M-H* loops over ± 800 Oe at *T* = 23.75 K and *T* = 24 K. Note first that the two loops are conspicuously shifted to the right and left of the origin, respectively. The *M-H* loop at an intermediate value of 23.85 K, shown in the inset panel (a) of Fig. 3 is, however, symmetric w.r.t. origin. The high field regions (± 5 kOe) of the *M-H* loops can be examined in the inset panel (b) in Fig.3, where the plots at 20 K, 23.75 K, 24 K and 27 K are shown. The *M-H* loops away from the compensation temperature at 20 K ($T < T_{comp}$) and 27 K ($T > T_{comp}$) show large hysteresis, whereas those closer to $T_{comp}$ at 23.75 K and 24 K are nearly collapsed. A considerable opening observed in the *M-H* loop at 20 K and 27 K is indicative of a large contribution from the uncompensated moments at these temperatures.

*Prima facie*, the behaviour of the alloy is that of a ferrimagnetic system at temperatures other than very close to $T_{comp}$, where a *quasi*-antiferromagnetism prevails and the *M-H* loop is nearly reversible. From the *M-H* loops at different temperatures, we estimated the effective coercive field (the half width of the loop, i.e., $H_c^{eff} = (H_+ - H_-)/2$, where $H_+$ and $H_-$ are the representative positive and negative field values where the net magnetization crosses $M = 0$ axis) and the exchange bias field (the shift in the *M-H* loop w.r.t. the origin, i.e. $H_{exch} = -(H_+ + H_-)/2$). The temperature variations of these two parameters are plotted in Fig. 4(a) and Fig. 4(b), respectively. The $H_c^{eff}(T)$ in Fig. 4(a) shows divergences in its temperature dependence as the $T_{comp}$ is approached from very low and high temperature ends. This behaviour undergoes a dramatic variation at temperatures very close to $T_{comp}$. *M-H* loops start to become asymmetric and their widths display shrinkage. $H_c^{eff}(T)$ 'dips' in the temperature interval 22 K to 26 K, where the exchange bias field surfaces as evidenced in Fig. 4(b). $H_{exch}(T)$ curve has a dispersion like shape, sign of the exchange bias field changes on going across $T = 23.85$ K, where $H_c^{eff}$ is minimum. In fact $H_c^{eff}$ remains in the ball park of its minimum value, while $H_{exch}(T)$ changes sign from maximum positive to minimum negative value.

The occurrence of positive exchange bias field (i.e., displaced hysteresis loop to the left of the origin) has been often discussed in the literature in terms of an exchange anisotropy at the interface of ferromagnetic/antiferromagnetic composites [14,15]. A schematic in Fig. 5 (a) illustrates the source of exchange anisotropy qualitatively in a multilayer structure [14]. Below $T_N$, the anisotropy in the interfacial coupling of a soft ferromagnetic layer with the two layers of an antiferromagnet can cause pinning of the ferromagnetic moments and result in the left shift of the hysteresis loop on the $M = 0$ axis. In another complex multilayer composite comprising notionally ferrimagnetic form of $GdCo_2$ and ferromagnetic Co, Webb *et al.* had reported [16] the divergence and the phase reversal in the exchange bias field across the magnetic compensation temperature of the $GdCo_2$ part. Webb *et al.* [17] had also reported (i) the divergence in the temperature dependence of the coercive field of the co-deposited amorphous form of Gd-Co film, which they termed as a microscopically homogeneous ferrimagnet and (ii) a collapse in the effective coercive field (as T → $T_{comp}$) of another Gd/Co multilayer composite, which they termed as inhomogeneous ferrimagnetic system. In the case of the admixed alloy being reported here, in the vicinity of $T_{comp}$, the magnetic contributions from both the rare earth moments are nearly compensated and the role of conduction electron polarization (CEP) becomes significant. One may be tempted to conjecture that CEP is an analogue of a soft ferromagnet of multilayer composite. Two schematics drawn in part (b) of Fig. 5 elucidate orientations of local moments of Nd/Ho and that of CEP w.r.t. the applied field (i) just above and (ii) just below $T_{comp}$. The similarity in the orientations of different sub-components in the latter case ($T < T_{comp}$) in part (b) of Fig. 5 and with that in its part (a) are apparent. However, the exchange bias field in the multilayer structure is typically positive in contrast to negative values observed at $T < T_{comp}$ in $Nd_{0.75}Ho_{0.25}Al_2$. It may be pertinent to reiterate that when divergence(s) were observed by Webb et al. [16] in the multilayered films of $GdCo_2$ and Co, the hysteresis loops were shifted to the right/left of the origin above/below their $T_{comp}$ of ~ 110 K, such a behavior is curiously phase reversed to the situation depicted in Fig. 4(b) for $Nd_{0.75}Ho_{0.25}Al_2$.

To rationalize the divergence and collapse in $H_c^{eff}(T)$, it is pertinent to recall a surmise made by Webb et al. [17] on the basis of their phenomenological model to comprehend the variety of behavior [17,18] in $H_c^{eff}(T)$ in Gd-Co composites. The occurrence of divergence(s) in $H_{eff}(T)$ in conventional ferrimagnets comprising two sublattices and displaying compensation behavior is well anticipated [17]. Webb et al. [17] had predicted that in ferrimagnets imbibing the notion of strong interfacial coupling between the two (distinct) components, the drop in effective coercive field would commence when temperature is close to $T_{comp}$. In one estimate, they projected this to happen for normalized $t$ $(= (T - T_{comp})/ T_{comp}) < 0.01$. If we view the data for $Nd_{0.75}Ho_{0.25}Al_2$ in Fig. 4(b) as $H_c^{eff}(t)$ versus t plot (see the top *x*-scale), the temperature interval ($\Delta t \sim 0.05$) of collapse appears to be of the right ball park. The stoichiometries in the admixed rare earth intermetallics, which are close to the no-net magnetization limit and display compensation behavior, are a special class of materials, which not only imbibe the characteristics of weak coupling between the two competing components of the phenomenological model of Webb et al. [17], but, they are also strongly coupled systems at the atomic level, as the two dissimilar rare earth moments occupy the same crystallographic site. The RKKY based exchange coupling operating between the spins of rare earth elements is expected to be long-ranged, thereby, assigning such systems the characteristic of homogeneous ferrimagnetic systems.

To summarize, we have presented highlights of our explorations of the magnetic compensation phenomenon in the single crystal of $Nd_{0.75}Ho_{0.25}Al_2$ alloy, which imbibes the notion of near-zero net magnetization. We have first shown that the characteristic features, like, the fingerprinting of field-induced reversal in the orientation of magnetic moments across magnetic compensation temperature and in the heat capacity and magneto-resistivity data, which have been described essentially in the context of Samarium based alloys in recent years [4-7, 10, 19], are infact generic to the stoichiometries in the admixed rare earth intermetallics, which are close to near-zero net magnetization. The interesting additional findings of the present work include, (i) identification of an exchange bias field and its sign change across $T_{comp}$, (ii) its correlation with the phase reversal in the CEP contribution to the net magnetization, (iii) collapse in the effective coercive field in close proximity of $T_{comp}$, exemplifying a prediction of the phenomenological description [15] given in the context of multi-layer composites, but, having wider validity.

**Acknowledgements**: We would like to gratefully acknowledge Mr. U. V. Vaidya and Mr. D. Buddhikot for their assistance in some of the measurements and Dr. D. Joshi for his help in cutting and orienting the crystals.

**Figure Captions**

Fig. 1. (Color online) The magnetization as a function of temperature in (a) nominal zero field and (b) 5 kOe in the three platelet shaped single crystal samples $Nd_{0.75}Ho_{0.25}Al_2$, for which the applied field could be easily oriented in three crystallographic directions. The nominal ordering temperature ($T_c$) and the magnetic compensation temperature ($T_{comp}$) are marked in the Fig. 1(a) at 70 K and 24 K, respectively. In Fig. 1(b), the magnetic turnaround temperature ($T^*$) has also been marked at 24 K.

Fig. 2. (Color online) The three main panels in Fig. 2 show the temperature dependences of (a) field cooled magnetization, (b) the specific heat and (c) the magnetoresistance in the single crystal of $Nd_{0.75}Ho_{0.25}Al_2$ for $H \parallel [100]$. In panel (a), the $M(T)$ curves in 10 and 20 kOe show that the turnaround temperatures vary with the applied field. The specific heat data in panel (b) shows a tiny peak close to 29 K (see inset (i) in Fig. 2(b) for data on an expanded scale), which corresponds to the field induced reorientation in Nd/Ho moments. The inset (ii) in Fig. 2(b) shows how the difference specific heat, $\Delta C_p$, scales with the external magnetic field. An inset panel in Fig. 2(c) displays the temperature dependence of the electrical resistance, in which the onset of a characteristic feature at $T_c$ can be easily marked. The magnetoresistance response in 10 kOe in Fig. 2(c) is oscillatory, with one of the sign reversals occurring at 29 K.

Fig. 3. (Color online) Magnetic hysteresis loops in a single crystal of $Nd_{0.75}Ho_{0.25}Al_2$ with $H \parallel [100]$. Main panel displays the portions of the *M-H* loops over ± 800 Oe at 23.75 K and 24 K, which are shifted to the left and right of origin, respectively. The inset panel (a) shows a portion of the *M-H* loop at 23.85 K, which is observed to be symmetric w.r.t. the origin. The inset panel (b) in Fig. 3 shows the *M-H* loops over ± 5 kOe at 20 K, 23.75 K, 24 K and 27 K. Note (i) the collapse in the width of the *M-H* loop near $T_{comp}$ values and (ii) the quasi-linear character in the *M-H* response at high fields ($H > 2$ kOe).

Fig. 4. Temperature dependences of (a) the effective coercive field, $H_c^{eff}$ (half-width of the *M-H* loop) and (b) the exchange bias field ($H_{exch}$) in [100] oriented single crystal of $Nd_{0.75}Ho_{0.25}Al_2$. Note the collapse of $H_c^{eff}(T)$ in the vicinity of $T_{comp}$. $H_{exch}(T)$ reverses sign in the narrow temperature range, where $H_c^{eff}(T)$ values are at a minimal level.

Fig. 5. Schematic representation of the orientation of applied field and those of different magnetic moments in (a) the multilayered structure, when the soft ferromagnetic layer is in contact with the two layers of an antiferromagnetic system at $T < T_N$, (b) the conduction electron polarization (CEP) exchange coupled to the two dissimilar RE moments ($\mu_{Nd}$ and $\mu_{Ho}$) at $T > T_{comp}$ and $T < T_{comp}$, respectively.

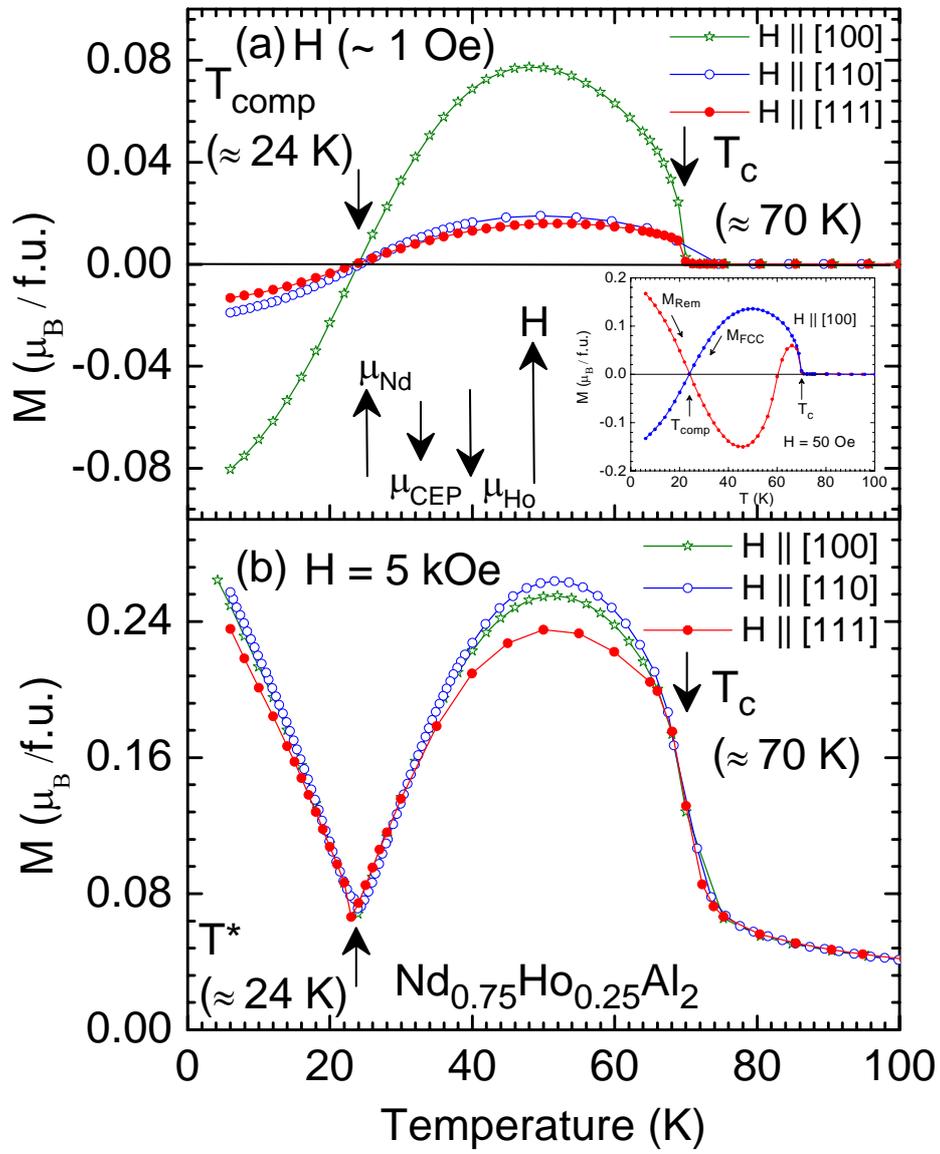

Fig. 1.

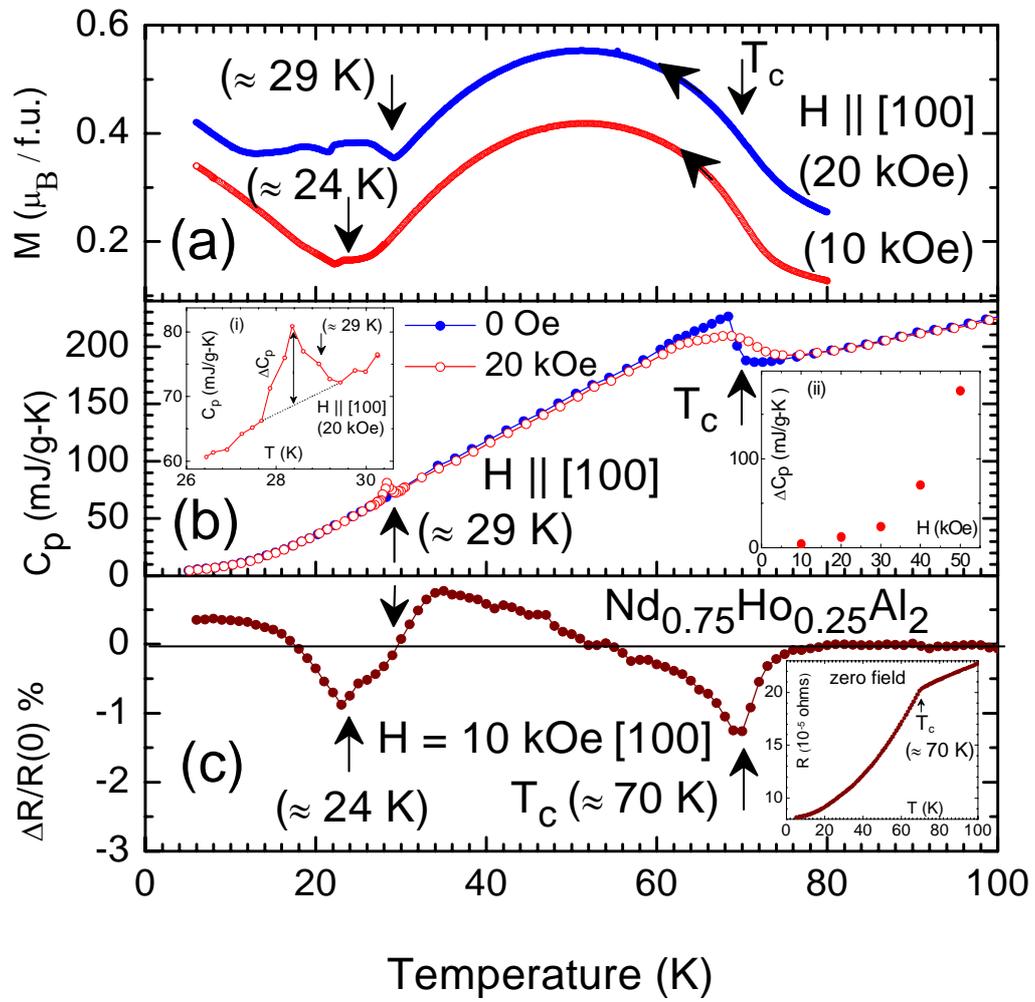

Fig. 2.

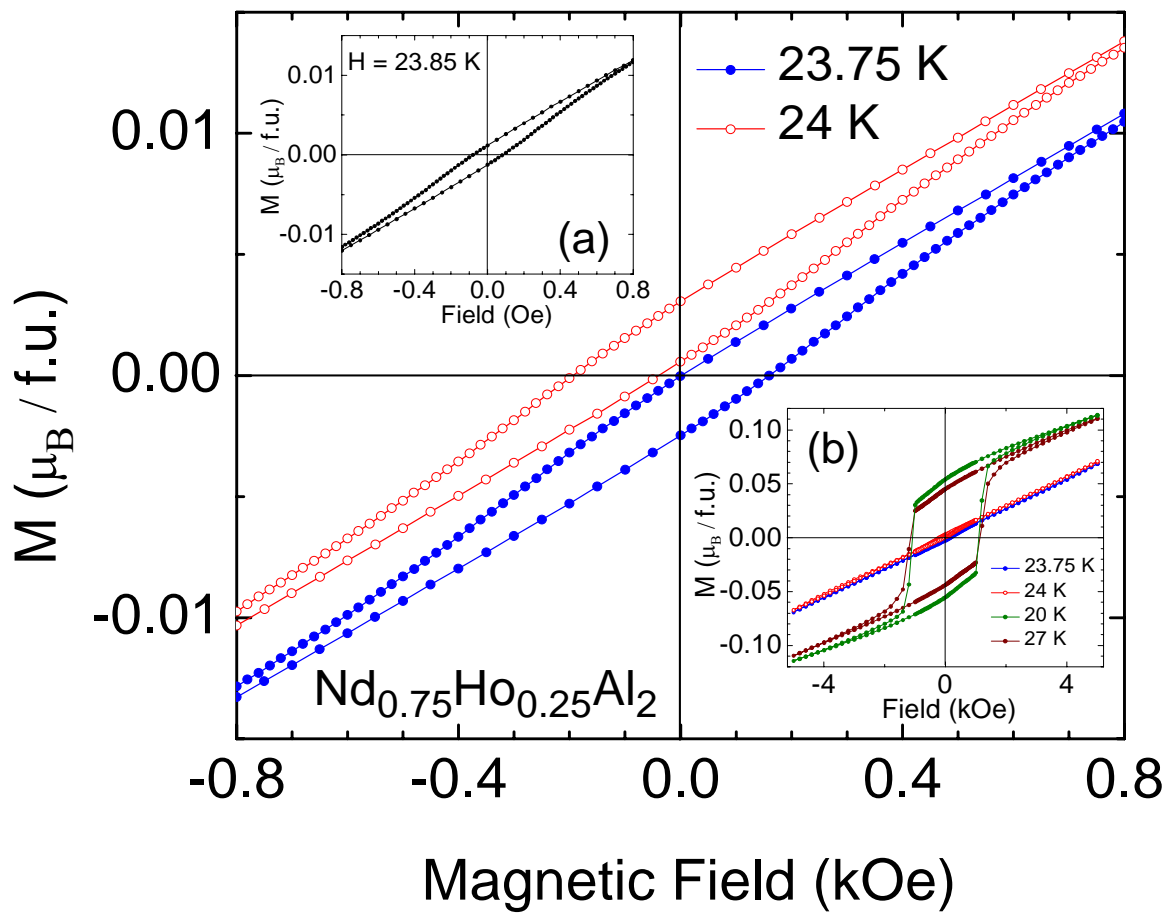

Fig. 3.

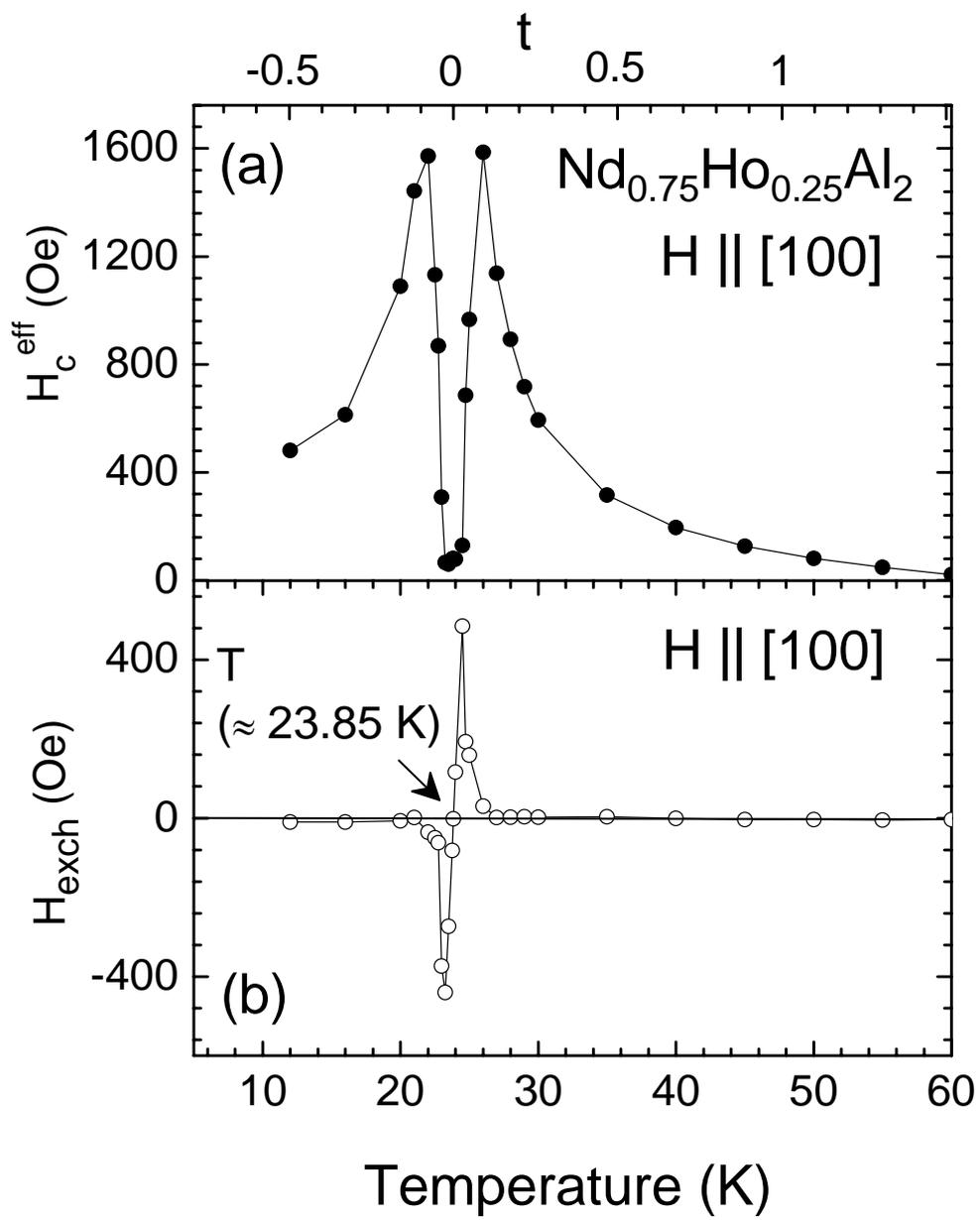

Fig. 4.

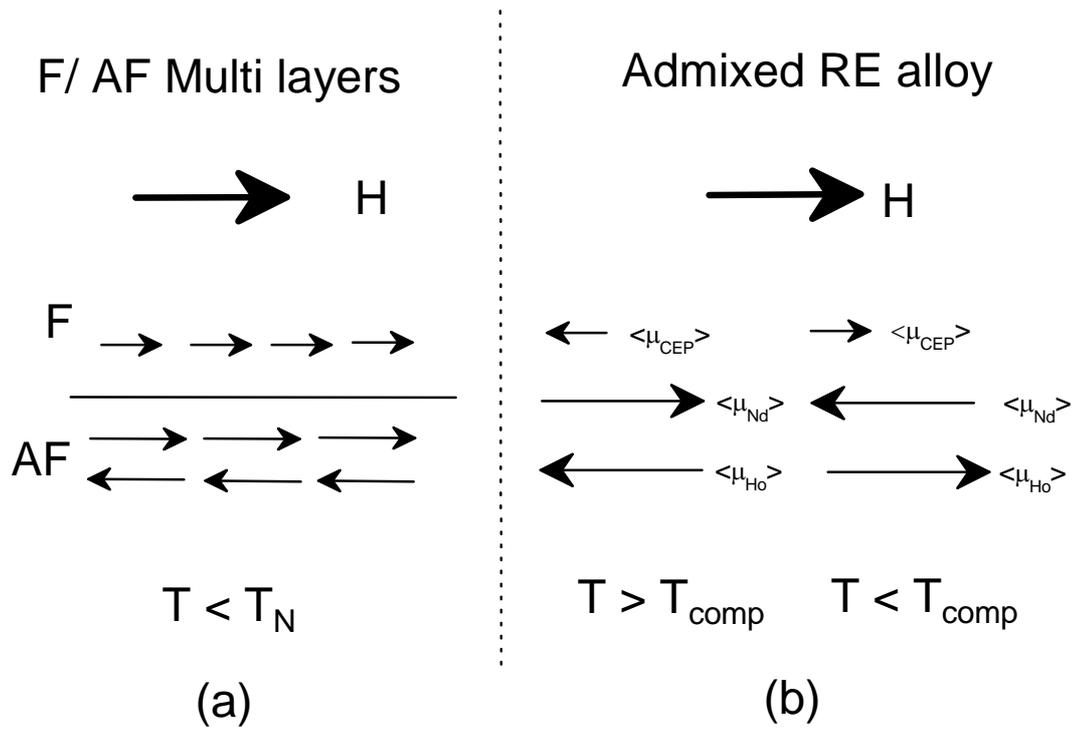

Fig. 5.